\documentclass[aps,pre,twocolumn,amssymb]{revtex4}
\usepackage[dvips]{graphicx}
\usepackage{dcolumn}
\usepackage{bm}
\def\gs{\gtrsim}

\def\be{\begin{equation}}
\def\en{\end{equation}}    
\def\gs{\gtrsim}

\newcommand{\bi}[1]{\mbox{\boldmath$#1$}}

\def\p{\partial}
\def\bea{\begin{eqnarray}}
\def\ena{\end{eqnarray}}

\begin{document}
\draft
\bibliographystyle{prsty}
\title{Critical adsorption  profiles  around a sphere and a cylinder  
in a  fluid at   criticality:\\ Local functional theory}
\author{Shunsuke Yabunaka$^a$  and Akira Onuki$^b$}
\address{
$^a$ Fukui Institute for Fundamental Chemistry, Kyoto University, Kyoto 606-8103, Japan \\
$^b$ Department of Physics, Kyoto University, Kyoto 606-8502,
Japan\\
}

\date{\today}

\begin{abstract} 
We study  universal critical adsorption 
on a solid sphere and  a solid cylinder in a fluid at bulk  criticality, 
where preferential  adsorption   occurs.  
 We use a  local functional theory 
  proposed by Fisher, de Gennes,    and Au-Yang 
($[$C. R. Acad. Sci. Paris Ser. B {\bf 287},  207 (1978)$]$ 
and $[$Physica {\bf 101}A,  255 (1980)$]$). We  calculate 
the mean order parameter profile $\psi(r)$, 
where $r$ is the distance from the sphere center and   the cylinder axis, 
respectively. The resultant  differential equation for  $\psi(r)$  
is  solved exactly around a sphere and numerically around a cylinder.  
A   strong adsorption  regime is  realized except  for  
very small surface field $h_1$, where 
 the surface order parameter $\psi(a)$ is determined by $h_1$ 
and is  independent of the radius  $a$.  
If $r$ considerably  exceeds  $a$, 
 $\psi(r)$  decays as  $r^{-(1+\eta)} $ for a sphere 
and $r^{-(1+\eta)/2} $ for  a cylinder in three  dimensions, 
where $\eta$ is the  critical exponent 
in the order parameter correlation  at bulk criticality.  
\end{abstract}


\maketitle

\pagestyle{empty}

\section{Introduction}

Surface critical phenomena and phase transitions  have 
long been studied in various near-critical systems 
\cite{PGreview,Cahn,Binder}. 
In particular,  the critical adsorption 
occurs when  a near-critical system  
in  a single-phase state is in contact with a 
distinctly different boundary \cite{Fisher-Yang,Fisher,PG,Naka}.  
In experiments, it has been studied intensively 
in near-critical  binary fluid mixtures 
on a solid surface or  a noncrtical vapor-liquid 
interface at constant pressure 
\cite{Desai,Be,Beysens,Franck,Fin,Cho,Lawreview,Liu}.  
Much attention has also been paid on 
preferential adsorption on colloidal particles 
\cite{Esteve,Omari,YLiu}, which 
is known to give rise to  colloid aggregation 
\cite{Esteve,Guo,Maher}.  We 
 mention a few experiments on critical adsorption of supercritical pure fluids  \cite{Chan,SF6,CO2}. 
There have been a large number of theoretical 
papers on the surface effects in the semi-infinite case 
\cite{Binder,Liu,Cahn,PGreview,Diehl,Bre,Upton-ad,Jasnow,Oka,Kie,Ci} 
and on curved surfaces \cite{Ei,PG,Hanke,Oka1,Ye}.

The order parameter in nearly incompressible binary mixtures 
is the deviation $\psi=c-c_c$ of the concentration 
$c$ from its critical value $c_c$.  
Because a  solid surface interacts with the two components differently, 
there can be  preferential adsorption on the surface. 
In such situations, the mean order parameter $\psi(z)$ 
 on a planar  surface 
(averaged over the thermal fluctuations) 
has been measured \cite{Beysens,Desai,Franck,Fin,Cho,Lawreview,Liu}, 
where $z$ is the distance from the  surface. 
From the Fisher-de Gennes scaling  theory \cite{Fisher}, 
it follows an algebraic  slow decay \cite{Naka,Jasnow,Liu,Upton-ad}, 
\be 
\psi(z)= Az^{-(1+\eta)/2}  ,
\en  
in three dimensions. Here,  $A$ is a constant and 
 $\eta$ is the  critical exponent $(\ll 1$) for 
 the correlation function of the critical fluctuations 
at bulk criticality. If   the bulk correlation length  
$\xi_{\rm B}$ is finite 
slightly away from the bulk  criticality, 
the above  form holds for $z< \xi_{\rm B}$ 
and $\psi(z)$ decays   exponentially as $\exp(-z/\xi_{\rm B})$ 
for larger $z$.

On the other hand, to examine the critical adsorption on a sphere, 
 de Gennes \cite{PG}  used a local functional 
theory by himself, Fisher, and Au-Yang \cite{Fisher,Fisher-Yang}
at bulk criticality.  He calculated  
  the   profile $\psi(r)$, where 
  $r$ is the distance from the  center of a sphere 
with radius $a$. Setting $\eta=0$, he found  that  
$\psi(r) $ largely drops in the range $a<r<2a$ 
and slowly decays as $ r^{-1}$ for  $r>2a$ 
in strong adsorption. With increasing $a$, this   strong adsorption condition 
can easily be realized even for  small  $h_1$. 
However, in their coarse-grained   free energy, 
   surface-dependence of the critical fluctuations is neglected 
 in   accounting   for  the renormalization effect. 
As a result, it  does not describe the  surface critical behaviors 
appearing at small surface field $h_1$ 
\cite{Binder,Diehl,Bre,Upton-ad,Jasnow,Ci,Cho}.  
Nevertheless,  their  free energy provides   a simple  
reasonable description of the critical adsorption 
in the  strong adsorption regime.

In this paper,  we examine the  adsorption  
on a sphere and a cylinder  at  bulk criticality  
with the local functional theory 
\cite{Fisher,Fisher-Yang}.  For a sphere, we find  an exact solution 
of the equation for $\psi(r)$ used  by de Gennes \cite{PG}. We  
 confirm his results and derive additional relations. 
For a cylinder, we examine it   numerically 
and  find  asymptotic behaviors of its  solutions. 
On  the other hand, in  colloidal systems with a near-critical solvent, 
 the thick-adsorption  regime  $a < \xi_{\rm B}$ can 
  be realized    for relatively small $a$  
 in the close vicinity of the critical point. 
However,   in this regime, we 
are aware of only one experimental report 
   by Omari {\it et al.}\cite{Omari}, so 
  its  physical   picture   remains  largely unexplored. 
Hence, this paper can be   a starting point 
in the research in this direction.

The organization of this paper is as follows. 
In Sec.II,  we will present   the   local functional theory 
 of  a critical fluid in the presence of a solid surface. 
In Sec.III, we examine 
the order parameter profiles 
near a planar wall, a sphere, and a cylinder. 
In Sec.IV, we will comment on 
the effect of motions of colloidal particles 
on the surrounding critical adsorption for $a< \xi_{\rm B}$. 

\section{Local functional theory}

We suppose  a solid surface  in a binary fluid mixture 
without ions, which is  
at its  critical point in the bulk at a given pressure.
The radius $a$ of the sphere or the cylinder 
 is much longer than the 
molecular length $a_0$($\sim 3~{\rm \AA}$ typically). 
The mean order parameter near a solid wall is denoted by $\psi$,  
which tends to $0$ far from the wall.

At bulk criticality, Fisher,  de Gennes,   
  and Au-Yang \cite{Fisher,Fisher-Yang}  
proposed  a local functional theory, 
where the  singular free energy consists of two terms as   
\be 
f= {B_0} |\psi|^{1+\delta}+ \frac{1}{2} C_0  |\psi|^{-\eta\nu/\beta}
 |\nabla \psi|^2 .
\en 
Here,  $B_0$ and $C_0$ are positive constants. 
The  spatial  variations of $\psi$ 
near the walls are  simply treated by the second gradient 
term. We use  the usual,  bulk critical exponents $\delta$, 
$\eta$, $\nu$, and  $\beta$   for Ising-like  
 systems. In three dimensions, we have 
$\beta\cong 0.33$,   $\nu\cong 0.630$, and 
 $\eta \cong 0.03-0.04$. 
For general space dimensionality $d$ ($2\le d\le 4)$,  
they satisfy the exponent  relations  \cite{Onukibook}, 
\be 
\delta= ({d+2-\eta})/({d-2+\eta}), 
 \quad 
{2\beta}/{\nu}= d-2+\eta.
\en

The ratio of the second term to the first term 
in $f$ can be  expressed 
as $\xi(\psi)^2|\nabla\psi|^2/4\psi^2$ \cite{Fisher,Fisher-Yang}, 
where 
\be 
\xi(\psi)= b_0 |\psi|^{-\nu/\beta}
\en 
is  the $\psi$-dependent  correlation length 
at $T=T_c$ with 
\be 
b_0= (2C_0/B_0)^{1/2} .
\en  
Here,  $b_0$ is a microscopic length 
if $\psi$ is the concentration deviation. 
This  $\xi(\psi)$  should be  longer 
than the molecular length $a_0$. 
The free energy in  Eq.(2)   is a renormalized one, 
where the thermal fluctuations 
shorter $\xi(\phi)$ have been coarse-grained, leading to 
 the fractional powers $|\psi|^{1+\delta -4}
$ and $|\psi|^{-\eta\nu/\beta}$ in $f$.  
As a result, it  cannot be used to describe   
 variations on scales shorter than $\xi(\phi)$.

From the two-scale factor universality 
\cite{HAHS,Stauffer,Onukibook},  the following combination 
is a universal number:
\be 
A_c= \xi(\psi)^d B_0 |\psi|^{1+\delta}/k_{\rm B}T_c= b_0^3 B_0/k_{\rm B}T_c.  
\en 
where we use $1+\delta= d \nu/\beta$. 
Then, $B_0=A_c k_{\rm B}T_c/b_0^3$ and  $C_0=A_c k_{\rm B}T_c/2b_0$.  
If  we  use the  $\epsilon$ expansion method 
of the renormalization group theory, 
 we  obtain  $A_c= 
18/(\pi^2\epsilon)+\cdots$ to first 
order in $\epsilon=4-d $ \cite{Oka,Okacomment}.

In the presence of a solid surface,  
the  free energy functional consists of bulk and surface 
parts as  \cite{Cahn} 
\be 
F= \int d{\bi r}f + \int d{S} f_s,
\en  
where the integral in the first term is 
performed in  the fluid  
and the second term is the surface integral of 
 the surface free energy  $f_s(\psi)$ 
 with $dS$ being the surface element. 
In this paper, we assume the linear form,  
\be 
f_s = -h_1 \psi,
\en 
where $h_1$ is  the surface field equal to   
the surface free energy difference between the two
components per unit area.    
We set  $h_1>0$ and have $\psi >0$. (For  $h_1<0$, 
we perform the sign changes:  $\psi \to -\psi$ and $h_1 \to -h_1$.) 
Assuming a significant  
size of $h_1$, we  neglect the second order term 
of the form $c \psi^2$  in $f_s$ \cite{Cahn,Binder}.

Following de Gennes \cite{PG}, we may   replace  
$1+\delta$ by 6 and  $ |\psi|^{-\eta\nu/\beta}$ by 1  
in $f$ setting  $d=3$  and  $\eta=0$.  
   Then,  minimization of $F$ yields 
the equilibrium conditions,  
\bea 
&&
\nabla^2 \psi= (6B_0/C_0)\psi^5  \quad ({\rm in~ fluid}),\\
&& {\bi n}\cdot\nabla \psi= -h_1/C_0\quad ({\rm on~surface}).
\ena 
where $\bi n$ is the outward normal unit vector on the surface. 
We also require  $\psi\to 0$ 
far from the solid surface 
where the  fluid is at  criticality.

To be precise,  we do not need  the above approximation 
($1+\delta \to 6$ and  $ |\psi|^{-\eta\nu/\beta} \to 1$) 
to obtain Eqs.(9) and (10). In fact, for general $d$ and nonvanishing $\eta$, 
 we introduce  the following variable 
 $\varphi$ by 
\be 
\varphi =  |\psi|^{-\eta/2\beta}\psi \quad {\rm or} \quad  
\psi= |\varphi|^{\eta_1}  \varphi .
\en  
where $\eta_1= \eta/(d-2)$. We then  find    $f= B_0|\varphi|^{2d/(d-2)} 
+ C_0 (1+\eta_1)^2|\nabla\varphi|^2/2$. See Appendix A for more details.

 Borjan and  Upton \cite{Upton-ad} calculated  $\psi(z)$ on a planar wall  
using a  local functional theory 
   in good agreement with results 
 of   Monte Carlo simulations.  Okamoto and one of the present authors (A.O.)
 \cite{Oka}  constructed  a  local functional theory in strong adsorption, 
which can be used  for  nonvanishing $T- T_c$ and $\psi$  
(including  the interior region of the coexistence curve). 
We could then study   
 phase separation between two parallel plates \cite{Yabu} and 
bridging between two colloidal particles  \cite{Yabu1,Oka1}. 

In semidilute  polymer solutions at the  theta condition, 
an order parameter   $\psi= \sqrt{\phi}$  
obeys the same equation as Eq.(9) 
\cite{PGpolymer}, where $\phi$ is the monomer volume fraction. 
This is because the free energy contains a gradient 
term  proportional to  $|\nabla\phi|^2/\phi$ 
and  no $\phi^2$-term at the theta condition.  
Thus, our results  can  also be used to describe  
polymer adsorption and depletion on solid walls (including colloid surfaces).

\section{Profiles near a planar wall, a sphere, and a cylinder}
\subsection{Planar surface}

We  first consider  $\psi(z)$ in the region $z>0$ on 
a planar surface at $z=0$, where    the $z$ axis is 
perpendicular to the surface. 
 Solving  Eq,(9), we find   
\be 
\psi(z)=  [b_0/4(z+ z_0)]^{1/2},
\en 
where   $z_0$ is a length determined by $h_1$ from Eq.(10) as 
\be 
z_0= b_0^{1/3}(C_0/4h_1)^{2/3}.
\en 
For general $d$ and   $\eta$, the exponent $1/2$ 
in Eq.(12) is changed to  $ \beta/\nu= 
(d-2+ \eta)/2$ (see Appendix A). Thus, Eq.(1) 
follows  for $z \gg z_0$ at  $d=3$.

The above  $z_0$   exceeds the microscopic length $a_0$  for 
\be 
b_0 h_1/C_0 \sim b_0^2 h_1/k_{\rm B}T_c < 1, 
\en 
where we assume $a_0\sim b_0$. 
Then, in the range $a_0 <z < z_0$, 
 $\psi$ assumes   the following surface value,  
\be 
\psi_s=(b_0/z_0)^{1/2}/2 = (b_0 h_1/2C_0)^{1/3},     
\en 
which is smaller than 1  under Eq.(14).
If the reverse inequality $b_0^2 h_1/k_{\rm B}T_c > 1 $  holds, 
the algebraic behavior (12) holds down to $a_0$ and 
the surface concentration  saturates to  1. 
For  $z_0\ll  \xi_{\rm B}$,  the  $z$ integral of $\psi(z)$ 
yields the excess adsorption 
 $\Gamma_{1d}\sim \xi_{\rm B}^{1-\beta/\nu}$ 
(independent  of $h_1$).

If we  define   the correlation length near  the surface 
by $\xi_s=\xi(\psi_s)=  b_0 \psi_s^{-2}$ from Eq.(4), we have  
$z_0=\xi_s/2  >  b_0 \sim a_0$ under Eq.(14). 
If we recover  $\eta$, 
we have the scaling relations 
$\xi_s \propto h_1^{-\nu/(2\nu-\beta)}$ 
and  $\psi_s \propto \xi_s^{-\beta/\nu}$. That is, 
with  Eqs.(2)  and (8) at  significant $h_1$, 
 there appear  no special surface critical exponents 
such as $\Delta_1$ and $\beta_1$  
\cite{Upton-ad,Jasnow}. 
However, for very small $h_1$, 
 it is known that 
the effect of the  surface 
on the critical fluctuations   becomes relevant  
  \cite{Diehl,Bre,Ci}.  
See the comment in Sec.I and 
 experiments in critical  mixtures at 
 very small $h_1$ \cite{Desai,Cho}.

\subsection{Sphere} 

We fix the position of   an isolated  solid sphere with radius $a$ 
in an near-critical fluid, where 
 $\psi$ depends only on the distance $r$ 
from the sphere center.  Experimentally, 
 the bulk correlation length $\xi_{\rm B}$  can much exceed  
$a$, where  our theory can be used in the 
space region $a<r<\xi_{\rm B}$.

\begin{figure}[t]
\begin{center}
\includegraphics[scale=0.55]{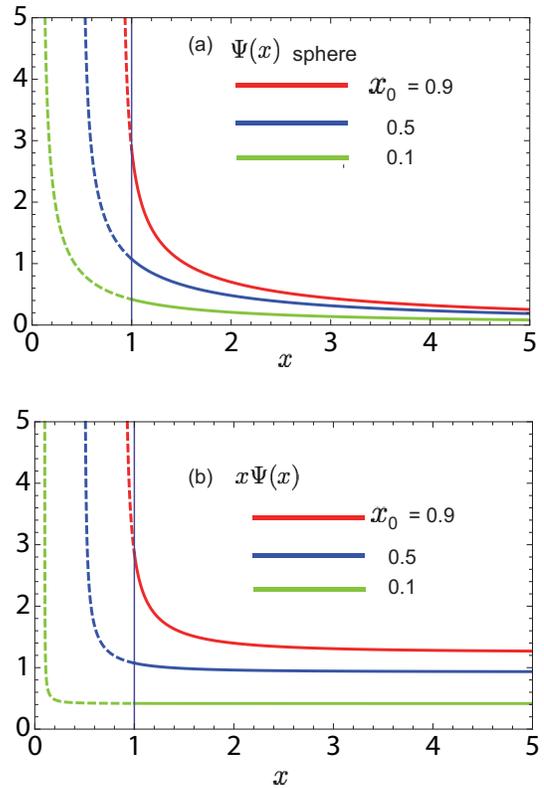}
\caption{\protect
(Color online) 
(a) Normalized order parameter   
$\Psi(x)$  and (b) $x\Psi(x)$ 
as functions of $x=r/a$  around a solid sphere 
with radius $a$ in a fluid at  criticality. 
They are written according to the exact solution (17) 
including the sphere interior ($x<1$). 
Three curves are obtained for $x_0= 0.9$, 0.5, and 0.1, 
for which $(H_1, \Psi_1,\alpha)=(15.1,2.86,1.64)$, $(1.43,1.08,1.22)$, 
and $(0.423,0.418,0.548)$, respectively. 
Here, $\Psi(x)$ diverges 
as $x\to x_0$ and decays as  $\alpha/x$ 
with $\alpha=3^{1/4}x_0^{1/2}$ 
for  $x>2$.  
}
\end{center}
\end{figure}

Starting with Eqs.(9) and (10) we  introduce a scaled order parameter $\Psi$ 
and a scaled surface field $H_1$ by 
\bea 
\Psi&=& 12^{1/4}b_0^{-1/2}a^{1/2} \psi, \\
H_1&=& 12^{1/4}b_0^{-1/2}a^{3/2}  h_1/C_0,  
\ena 
where $H_1= 3^{1/4}(a/2z_0)^{3/2}$  in terms $z_0$ in Eq.(13)\cite{PG}.
Using  the scaled distance $x =r/a$ we obtain 
 \bea 
&&\Psi'' + 2x^{-1}\Psi' = \Psi^5,\\
&& \Psi'= -H_1 \quad (x=1) ,
\ena  
with  $\Psi(\infty)=0$.  
Here,  $\Psi'= d\Psi/dx$ and  $\Psi'' = d^2\Psi/dx^2$.

 Remarkably, Eq.(18) can   be solved exactly as 
\be 
\Psi(x)= 3^{1/4} x_0^{1/2}/({x^2- x_0^2})^{1/2}, 
\en 
which diverges as $x\to x_0$ with   $x_0$ 
being a lower bound  in the range $[0,1]$ (inside the sphere). 
In terms of $x_0$, $H_1$ and $\Psi_1= \Psi(1)$ are expressed as  
\be 
H_1 =\Psi_1 /({1- x_0^2}) = 3^{1/4} x_0^{1/2}/({1- x_0^2})^{3/2}.  
\en 
We plot $\Psi(x)$ and  $x\Psi(x)$  
in  Fig.1,  while   we show  the 
relations among $x_0$, $H_1$,  $\Psi_1$ and $\alpha$ in  Fig.2 
(see Eq.(26) for $\alpha$). In  Appendix A, we shall see   
 the exact profile  around a sphere  at bulk criticality 
for general $d$ and  $\eta$ using Eq.(3). 
See Remark (i) in Sec.V   also. 

We are interested in the   strong adsorption  
regime $H_1 \gg 1$, 
which is realized  for $b_0 h_1/C_0 \gg (b_0/a)^{3/2}$ 
from  Eq.(17). For large  $a$, 
this condition can be  realized even under Eq.(14). 
 From Eq.(21) we express $x_0$ and $\Psi_1$ in terms of $H_1$ in the 
 weak and strong adsorption limits as    
\bea 
&& \hspace{-5mm} 
x_0 \cong 3^{-1/2} H_1^2, \quad  \Psi_1 \cong H_1 ~\quad (H_1 \ll 1),
 \\
&& \hspace{-1cm} 1 -x_0^2 \cong 
{3^{1/6}} H_1^{-2/3},\quad \Psi_1\cong 3^{1/6}H_1^{1/3}
  \quad (H_1 \gg 1).
\ena 
For $H_1\gg 1$, $x_0$ approaches 1 and $\Psi_1$ grows. 
From Eqs.(16) and (21)-(23)   the  surface order parameter $\psi_s$ 
in the original units behaves as follows: 
\bea 
\psi_s &=& \psi(a) 
\cong  ah_1/C_0 \quad (H_1\ll 1) \\
&\cong&   (b_0 h_1/2C_0)^{1/3} \quad (H_1 \gg 1)
\ena  
We can derive Eq.(24)  
 from $\psi\cong \psi_s a/r$ ($r>a$) for $H_1 \ll 1$. 
Notice that Eq.(25) 
is of the same form as  Eq.(15) for a planar wall. 
See Appendix A for this point.

\begin{figure}[t]
\begin{center}
\includegraphics[scale=0.42]{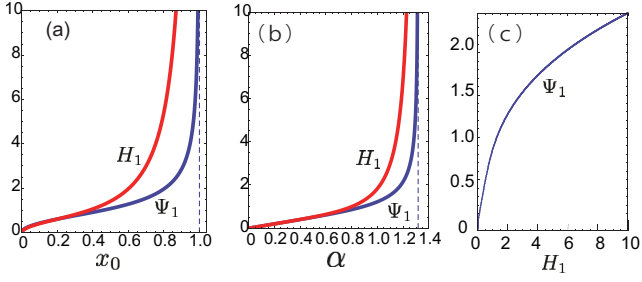}
\caption{\protect
(Color online) (a) $H_1$ and $\Psi_1$ vs $x_0$ 
from Eq.(21), where $0<x_0<1$. (b)  $H_1$ and $\Psi_1$ vs 
$\alpha= 3^{1/4} x_0^{1/2} (<3^{1/4})$. 
(c)  $\Psi_1$ vs  $H_1$, which behaves 
as in  Eqs.(22) and (23). 
}
\end{center}
\end{figure}

For  $x \gg x_0$, we find 
 the slow  decay $\Psi \cong \alpha/x$ with \cite{PG} 
\be 
\alpha= 3^{1/4}x_0^{1/2}. 
\en 
This  $\alpha$  should not be confused with the critical 
exponent $\alpha$.    
In the original units, this  decay  is rewritten as
\be 
\psi (r) \cong (x_0b_0 /2)^{1/2}a^{1/2} r^{-1}.
\en 
For $H_1 \ll 1$, 
the above  behavior holds in the whole fluid, so $\psi(r) \cong \psi_s a/r$ 
for   $r>a$ (see Fig.1).  For  $H_1  \gg  1$,  it  holds for 
$r/a$ considerably larger than 1, say, for $x= r/a\gs 2$, 
where the correlation length  in Eq.(4) is 
estimated as 
\be 
\xi(\psi)\sim r^2/a\quad (r\gs 2a). 
\en 
If $\xi_{\rm B}$ is finite, the above relation 
is meaningful for $r< (a\xi_{\rm B})^{1/2}$. 
In this case, from Eqs(20) and (23), 
we calculate the drop of $\psi(r)$ 
in the region $1<r/a<2$ as  
\be 
\psi(2a)/\psi(a)= \Psi(2)/\Psi(1) 
\sim (1-x_0^2)^{1/2} \sim H_1^{-1/3},
\en 
which is  smaller than 1 for $H_1\gg 1$. 
If $\eta$ is recovered, 
Eq.(11) leads to 
$\psi \sim  (b_0/a)^{(1+\eta)/2}\Psi^{1+\eta}$  so that 
\be 
\psi (r)  \sim (x_0 b_0 a)^{(1+\eta)/2} r^{-(1+\eta)}.
\en  
See Eq.(A4) in  Appendix A. We may 
also consider  the total excess adsorption, written as 
$\Gamma_{\rm tot}$,  which is  
the space integral of $\psi$ outside the sphere. 
For    $H_1 \gg 1$ and   $a\ll \xi_{\rm B}$, 
the contribution from  the region $r>2a$ 
is dominant and is estimated from Eq.(30) as \cite{Hanke}
\be  
\Gamma_{\rm tot} 
\sim a^{(1+\eta)/2}\xi_{\rm B}^{2-\eta}.
\en

It is worth noting  that the pair correlation function of the 
critical  fluctuations of $\psi$ 
decays  at criticality  as  \cite{Onukibook} 
\be 
g_{\rm th}(r)= C_{\rm th} r^{-(d-2+\eta)}.
\en  
Setting   $d=3$ and $\eta=0$, we find 
 $C_{\rm th}\cong k_{\rm B}T_c/C_0=  2b_0/A_c$ in terms 
of  $C_0$  in Eq.(2). Interestingly, 
$\psi(r)$ in Eq.(30)   depends on $r$ 
in  the same manner as  $g_{\rm th}(r)$.   
The coefficient in  the former  grows  as $a^{(1+\eta)/2}$ 
 with increasing $a$ in the strong adsorption regime.

Originally,  de Gennes \cite{PG} numerically calculated 
 a  special solution of Eq.(18), 
written as $\Phi_{\rm PG}(x)$, which behaves as 
 $x^{-1}$ for  $x \gg 1$. 
He constructed the other solutions by 
 scaling $\Phi_\alpha(x)= \Phi_{\rm PG}(x/\alpha^2)/\alpha$ 
(see Eq.(36)), 
which tends to  $\alpha/x$ for large $x$. 
In our theory,    Eq.(20) gives   
$ 
\Phi_{\rm PG}(x)=(x^2-1/3)^{-1/2} 
$  with  $x_0= 3^{-1/2}\cong 0.58$. 
 Burkhardtt and  Eisenriegler \cite{Ei} 
found the  profile in 
Eq.(20) particularly for $x_0=1$ (where $H_1=\infty$)  
using conformal mapping of the results for the half space.


\subsection{Cylinder}

We next consider a cylindrical wire fixed in a critical fluid. 
It is infinitely elongated and  
 $\psi$ depends only on the distance $r$ from the cylindrical axis.  
We use the normalized $\Psi$ and $H_1$ in Eqs.(16) and (17). 
In terms of $x=r/a$, Eq.(9) becomes 
 \be
\Psi'' + x^{-1}\Psi' = \Psi^5 . 
\en 
The boundary condition at $x=1$ is given by Eq.(19) and we  
assume $\Psi(\infty)=0$.  In particular, for $H_1 =2^{-3/2}\cong 0.354$,
we find   a special solution  given by 
\be 
\Psi_{\rm sp}(x) =(2x)^{-1/2} .
\en  
In the original units, Eq.(16) gives   
\be 
\psi_{\rm sp} 
= 12^{-1/4} (b_0/a)^{1/2}\Psi_{\rm sp}
= 3^{-1/4} (b_0/4r)^{1/2}, 
\en  
which is independent of $a$. If $r$ here is replaced by $z+z_0$, 
 this solution is smaller than the one-dimensional 
profile (12) by $3^{-1/4}$. 
For general $d$ and   $\eta$, we obtain $\psi_{\rm sp} \propto 
r^{-(d-2+\eta)/2}$ (see Appendix A). 
However,   we can solve Eq.(33)  only numerically  for general $H_1$. 
In   Fig.3, we plot  $\Psi(x)$ and  $(2x)^{1/2}\Psi(x)$ 
for three values of $H_1$, where the latter 
behaves differently for positive and negative 
$H_1 - 2^{-3/2}$. 
In   Fig.4, we display $\Psi_1$ vs $H_1$ 
for these two cases separately.

\begin{figure}[t]
\begin{center}
\includegraphics[scale=0.55]{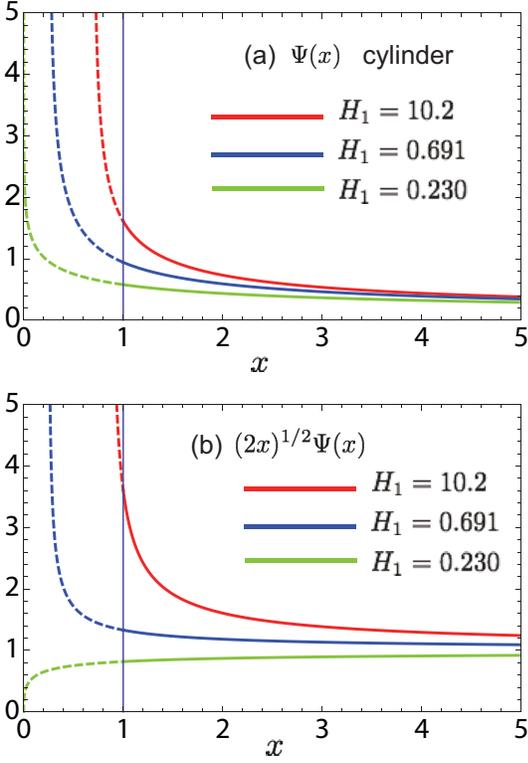}
\caption{\protect
(Color online) 
(a) Normalized order parameter   
$\Psi(x)$  and (b) $\sqrt{2x}\Psi(x)$ 
as functions of $x=r/a$  around a solid  cylinder  
with radius $a$ in a fluid at  criticality. 
They are numerically obtained from Eqs.(33) and (19) 
for $H_1= 10.2$, 0.691, 
 and 0.230, for which $ \Psi_1=1.61$, $0.941$, and $0.579$, 
 respectively. They  are written in  the cylinder exterior and 
 interior. For $H_1> 2^{-3/2}=0.354$, $\Psi(x)$ diverges as $x\to x_0$ 
as in  Eq.(40), where $x_0$ is 0.871 for $H_1=10.2$ 
and is 0.248 for $H_1= 0.691$. 
For $H_1< 2^{-3/2}$,  $\Psi(x)$ diverges 
  as $x\to 0$ as in  Eq.(41). 
In (b) $\sqrt{2x}\Psi(x)$    tends to   $1$ at  large $x$ 
as in Eq.(38)  for any $H_1$. 
}
\end{center}
\end{figure}

\begin{figure}[t]
\begin{center}
\includegraphics[scale=0.46]{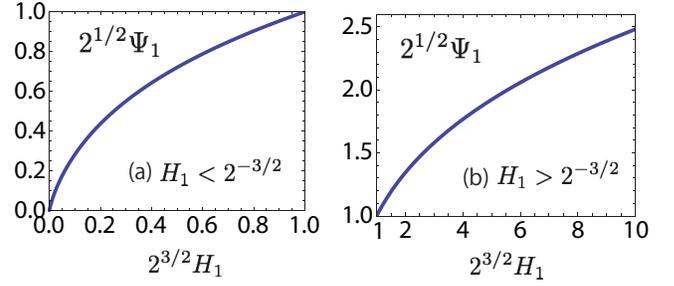}
\caption{\protect
(Color online)  $2^{1/2} \Phi_1$ vs  $2^{3/2} H_1$ for a cylinder 
for  (a) $H_1< 2^{-3/2}$ 
and  (b)  $H_1>2^{-3/2}$.  These curves are calculated 
from Eq.(37) using two special solutions in these cases. 
The curves behaves as in Eqs.(42) and (43). 
}
\end{center}
\end{figure}

Let  $\Psi_0(x)$ be  a (numerically calculated) 
special solution 
of  Eq.(33) with $H_1 \neq 2^{-3/2}$. 
Then, other solutions of Eq.(33) 
can be obtained by de Gennes'  scaling \cite{PG}:  
\be 
\Psi_\lambda(x)= \lambda^{1/2}\Psi_0(\lambda x) 
\en 
where $\alpha$ in Ref.\cite{PG} is replaced by $\lambda^{-1/2}$. 
For $\lambda<1$  we need to know the behavior of $\Psi_0(x)$ 
in the range $x<1$.   We can then relate 
 $\Psi_1=\Psi(1)$ and $H_1$ by eliminating $\lambda$ 
in  the following relations: 
\be 
\Psi_1= \lambda^{1/2} \Psi_0 (\lambda), \quad 
H_1= -\lambda^{3/2} \Psi_0'(\lambda) ,
\en

From analysis in Appendix B, we  further find the following. 
(i) For any $H_1$,  $\Psi$ behaves for $x \gg 1$ as  
\be 
\Psi (x) =   (2x)^{-1/2}[ 1+ G_1  x^{-\zeta}+\cdots],
\en 
where $G_1$ is a constant and  $\zeta=(\sqrt{5}-1)/2$. 
Thus, $\Psi (x) \to \Psi_{\rm sp}(x)$ slowly  for any $H_1$, 
which is consistent with  the slow decay of  $(2x)^{1/2}\Psi(x)$  in Fig.3(b). 
In particular, for small $H_1 -2^{-3/2}$, we obtain the linear relations, 
\be 
G_1 \cong 5^{-1/2}(2^{3/2}H_1-1)\cong 2^{1/2}\Psi_1-1. 
\en 
The excess adsorption per unit length is 
of order $\xi_{\rm B}^{(3-\eta)/2}$  and is independent of 
$a$ for  $a\ll \xi_{\rm B}$.

(ii) If  $H_1> 2^{-3/2}$,  $\Psi(x)$ is larger than  $(2x)^{-1/2}$ for any $x
>x_0$ with $0<x_0<1$.  It  diverges as $x\to x_0$ as  
\be 
\Psi(x) \cong 3^{1/4} [2x \ln (x/x_0)]^{-1/2}.
\en 
For example,  we have 
 $x_0= 0.248$  for $H_1= 0.691$. 
Use of Eq.(36) yields other solutions diverging 
at $\lambda^{-1}x_0$ \cite{comment2}. 

(iii)  If  $H_1< 2^{-3/2}$, we have 
$\Psi(x)<(2x)^{-1/2}$ for any $x>0$.  As  $x \to 0$,  we find   
\be 
 \Psi(x) \cong  2^{-1/2}(A_1- A_2 \ln x) ,
\en 
 where $A_1$ and $A_2$ are constants. 
For example,  $A_1= -1.27$ and $A_2= 1.34$ 
for $H_1= 0.3$ \cite{comment2}.

Now, in   Eq.(37),  we  set 
 $\Psi_0(\lambda)= 3^{1/4} [2\lambda \ln (\lambda/x_0)]^{-1/2}$ from Eq.(40) 
and  $\Psi_0(\lambda)= 2^{-1/2}(A_1- A_2 \ln \lambda)$ 
from Eq.(41). Here,  the former (latter) 
yields the limiting behavior for large (small) $H_1$ as follows: 
\bea 
\Psi_1 &\cong& 
H_1 (A_3 -2\ln H_1 )  \quad (H_1\ll 1),\\
 &\cong& 3^{1/6} H_1^{1/3} \quad (H_1\gg 1)
\ena
where $A_3=A_1/A_2+ \ln(A_2^2/2)\cong -1.05$ \cite{comment2}.
Notice that  Eq.(43) coincides with the second relation in Eq.(23), 
 so the surface order 
parameter $\psi_s$  is again 
given by  Eq.(15) or Eq.(25) in the strong adsorption regime. 
In fact, from Eq.(40), we 
find  $\Psi(x)\cong (3/4)^{1/4}/(x-x_0)^{1/2}$ 
for $x_0\cong 1$ and $0<x-1\ll 1$ 
as in the sphere  case (see the remark below Eq.(25)).

\section{Adsorption around moving colloidal particles}

So far we have fixed the position of a sphere 
or a cylinder. We may also 
 suppose   colloidal particles in a near-critical fluid 
\cite{Esteve,Guo,Maher,Omari,YLiu}. 
Beysens' group \cite{Esteve} observed an increase of the scattered 
 light intensity due to formation of  adsorption layers 
on colloid surfaces \cite{Esteve}, where $ \xi_{\rm B}/a \sim 0.1$ typically in their experiments. 
Such thin layers should remain  attached to the particles   
during their thermal Browninan   motions. However, 
 for $a< \xi_{\rm B}$, it is not clear how the adsorption 
occurs far from the surfaces.
 Here, we  argue that 
the particle motions should prevent 
establishment of long-ranged  absorption profiles.

We consider a colloidal particle with radius $a$ 
in the strong adsorption condition $H_1\gg 1$, where the fluid is at  
criticality far from it. 
In this case, we can introduce  a space-dependent 
 relaxation time $t_\xi$ of 
the critical fluctuations 
around the particle by  \cite{Onukibook}
\be 
t_\xi= 6\pi \eta_s \xi(\psi)^3 /k_{\rm B}T_c,
\en   
where  $\xi(\psi)$  is 
the local correlation length 
in  Eq.(4) depending on $r$ as in Eq.(28). The   $\eta_s$ 
is the viscosity, which 
may be treated as a constant due to its weak  singularity. 
On the other hand, the  particle  undergoes Brownian  motions, 
   moving  a distance of $a$ 
 on the diffusion  time,  
\be 
t_a= a^2/D= 6\pi \eta_s a^3 k_{\rm B}T_c,
\en 
where $D=k_{\rm B}T_c/6\pi\eta_s a$ is the  diffusion constant. 
From Eq.(28),  the ratio 
 $t_a/ t_\xi$ is larger than 1   for $a<r<2a$ 
 but is smaller than 1  for $r>2a$. Hence, by the  diffusion, 
the profile  $\psi(r)$ should not be 
affected   in the   vicinity  ($a<r <  2a$), 
but it should be largely deformed    far from 
the particle  ($2a<r<\xi_{\rm B})$. 
As a result, for $a<\xi_{\rm B}$, 
the   excess  adsorption  of  a diffusing particle 
should be significantly smaller than that of a fixed particle  in Eq.(31).

As a related   experiment,  
 Omari {\it et al.} \cite{Omari} 
 determined  the hydrodynamic radius  $R_{\rm H}$ of 
colloidal particles with  $a=25$ and $10$ nm 
from their diffusion constant. 
Remarkably, the effective layer thickness   $R_{\rm H}-a$ 
increased above $a$  on approaching the critical point, 
where the maximum of  $R_{\rm H}$ was about $5a$  
for $\xi_{\rm B}\sim 80$ nm. As argued above, the diffuse  boundaries of 
such thick layers should be  nonstationary and nonspherical. 
For not very small colloid densities,   
the interaction among the particles 
 also appears with increasing the  layer thickness.   
 Further  experiments   are  informative 
to understand  these aspects.

We mention    simulations 
on  the dynamics of colloidal particles with  adsorption  layers, which  
include the hydrodynamic interaction. Furukawa {\it et al.} \cite{Furukawa} 
 found deformations of   thick  adsorption layers   
at the critical composition. Yabunaka {\it et al.} \cite{Yabu1} examined 
the  bridging dynamics between two particles at off-critical compositions using the local functional theory, where the adsorption 
layer remained  attached to the surfaces 
for $\xi_{\rm B}\sim 0.1a$. Barbot and Araki \cite{Araki} 
studied   aggregation and rheology  of a large number of  colloidal particles 
outside the solvent coexistence curve. However, we need  further simulations 
 accounting for  the Brownian motions 
 in the case $a<\xi_{\rm B}$.

\section{Summary and remarks}

We have calculated the order parameter profile $\psi(r)$  
around a sphere and a cylinder 
 fixed  in a fluid at bulk  criticality, where 
the radius $a$ is  longer than the microscopic length 
$a_0$. Following de Gennes \cite{PG}, we have  used   the 
 local free energy in Eq.(2)  
and the surface free energy in Eq.(8) with 
significant surface field $h_1$.

In Sec.III,  setting  $\eta=0$ and $d=3$, 
we have solved de Gennes' equation to find the following. 
(i)  The strong adsorption regime is realized 
when the normalized surface field $H_1 (\propto a^{3/2}h_1$) 
exceeds 1.  (ii) For $H_1 \gg 1$, the surface order parameter 
$\psi_s =\psi(a)$  grows up to a value ($\propto h_1^{1/3}$) 
independent of $a$, which  
coincides with the  one  on a planar surface. 
 (iii) We have found  the  exact profile 
$\psi \propto  (r^{2}- x_0^2a^2)^{-1/2}$ 
 ($0<x_0<1$)  around a sphere   
and the asymptotic decay  
$\psi \to  \psi_{\rm sp}\propto  r^{-1/2}$ ($r\gg a)$  around a cylinder. 
In Appendix A, these expressions 
become   $\psi \propto (r^{2}- x_0^2a^2)^{-\beta/\nu}$ 
  around a sphere  
and  $\psi \propto r^{-\beta/\nu}$  
  around a cylinder with 
$\beta/\nu= (d-2+\eta)/2$ for 
 general $d$ and nonvanishing $\eta$.

In Sec.IV, we have argued that the  Brownian 
motions of colloidal particles strongly affect their  
thick  adsorption layers for  $a<\xi_{\rm B}$. 
The physics in this case   has not been examined in the literature. 
We need to fix a solid sphere or a  cylinder  in space 
to confirm the predicted critical long-range adsorption.

We further make critical remarks  as follows. 
(i) The long-range decay $\psi(r) \propto r^{-(d-2+\eta)}$ 
around a sphere is of the same form as the correlation function 
of the order parameter fluctuations at bulk criticality, 
as has been discussed around Eq.(32). 
Note that the equation $\Psi''+ (d-1)x^{-1}\Psi'= \Psi^\lambda$ 
around a sphere ($x=r/a$) 
can be solved  exactly  for $\lambda=(d+2)/(d-2)$ 
as in Eq.(A4).  In particular, at the mean 
field criticality in three dimensions 
($d=\lambda=3$),  we find  
$\Psi \cong x^{-1} (2\ln x + A_m)^{-1/2}$ 
 for $x\gg 1$  ($A_m$ being  a  constant) 
and  $\Psi \propto 
(x- x_0)^{-1}$ as $x\to x_0$ ($x_0<1$). 
(ii)  Though we have 
calculated $\psi$ at the critical composition 
 (for mixture solvents), 
the preferential adsorption 
is much  enhanced 
  when the solvent component favored by 
the surface is poorer than the other one 
  in the bulk \cite{Oka,Oka1,Yabu1}.
This off-critical enhancement  is crucial 
 in the observed phenomenon of   colloid aggregation 
\cite{Esteve,Maher,Guo}. 
In this case    $\psi(r)$ 
passes through zero at $r-a \sim  \xi_{\rm B}$ 
since $\psi(a)>0$ and $\psi(\infty)<0$. 
(iii) As stressed in Sec.IV, further simulations are needed to investigate 
the solvent-mediated  colloid interactions  
  in the case $a<\xi_{\rm B}$.

\begin{acknowledgments}
This work was supported by KAKENHI 15K05256. 
A.O. would like to thank D. Beysens for informative correspondence. 
S.Y. was supported by Grant-in-Aid for Young  Scientists
(B) (15K17737), Grants-in-Aid for Japan Society for 
Promotion of Science (JSPS) 
Fellows (Grants No. 263111), and the JSPS Core-to-Core Program 
''Non-equilibrium dynamics of soft matter and information''.
\end{acknowledgments}

\vspace{2mm}
\noindent{\bf Appendix A: Profiles for general $d$ and nonvanishing 
$\eta$ at bulk criticality  }\\
\setcounter{equation}{0}
\renewcommand{\theequation}{A\arabic{equation}}

We seek the exact profiles minimizing $F$ in Eq.(2) 
for general $d$ and nonvanishing $\eta$ using Eq.(3). 
In terms of $\varphi$ in Eq.(11), we obtain 
the bulk equilibrium relation,   
\be 
(\beta b_0/\nu)^2 \nabla^2 \varphi= d(d-2) \varphi^{(d+2)/(d-2)},
\en 
and the boundary condition on the surface,   
\be 
{\bi n}\cdot\nabla \varphi= -(1+\eta_1)^{-1} C_0^{-1} \varphi^{\eta_1} h_1 .
\en 
Here, we have used Eqs.(3) and (5) with  $\eta_1=\eta/(d-2)$.

 First, let the one-dimensional solution  of Eq.(A1) be 
written as $\varphi_{\rm p} (z)$, 
which behaves as $(z+z_0)^{-1/2}$ with 
 $z_0$ being  a positive constant.  From Eq.(11), 
 the order parameter profile $\psi_{\rm p}(z)$ is written as  
\be 
\psi_{\rm p}(z)=\varphi_{\rm p}(z)^{1+\eta_1}= 
 [\beta b_0/2\nu(z+z_0)]^{\beta/\nu}. 
\en

 Second, for a 
sphere with radius $a$, we set 
 $\nabla^2= d^2/dr^2+ (d-1)r^{-1}d/dr$ in Eq.(A1) to obtain 
$\varphi_{\rm s}(r) \propto (r^2/a^2-x_0^2)^{-1/2}$. 
The    profile $\psi_{\rm s}(z)$ is expressed  as 
\be 
\psi_{\rm s}(r)=\varphi_{\rm s}(r)^{1+\eta_1}= 
 [\beta b_0 x_0a /\nu(r^2- x_0^2a^2)]^{\beta/\nu},
\en 
where $x_0$ is in the range $[0,1]$. 
Here, in  the limits  $x_0\to 1$ and $r/a-1 \ll  1$,  
 Eq.(A4) leads to Eq.(A3) with  replacement  $r-x_0a \to z+ z_0$ 
with $z_0= a(1-x_0)$.

Third, for a cylinder with radius $a$, we set 
 $\nabla^2= d^2/dr^2+ (d-2)r^{-1}d/dr$.
In this case, we can find  an exact  solution 
$\varphi_{\rm sp}(r)$ only for a special value of $h_1$ 
as in Sec.IIIC (where $d=3$ and $\eta=0$). 
It is expressed as $\varphi_{\rm sp}(r)\propto r^{-(d-2)/2}$  so that 
\be 
\psi_{\rm sp}(r)=\varphi_{\rm sp}(r)^{1+\eta_1}\propto  
  r^{-\beta/\nu}.
\en   
Around a cylinder, the  profile  $\psi (r)$   
tends to $\psi_{\rm sp}(r)$ at 
 large $r\gg a$ for any $h_1$, as shown in Sec.IIIC.

\vspace{2mm}
\noindent{\bf Appendix B: Profile around a cylinder}\\
\setcounter{equation}{0}
\renewcommand{\theequation}{B\arabic{equation}}

We derive the behaviors of $\Psi(x)$ in Fig.3 
in  the case of a cylindrical wire.   To this end, we rewrite 
Eq.(33) in terms of $w(x) = (2x)^{1/2}\Psi(x)$ as 
\be 
4x^2 w''= w^5-w ,
\en 
where $w''= d^2 w/dx^2$. This  is surely satisfied for $w=1$ 
as a special solution with $H_1= 2^{-3/2}$. 
For any $H_1$, $w$ tends to 1 at large $x$.  
If we  linearlize Eq.(B1)  with respect to the deviation  $ w_1 =w-1$, we 
obtain $x^2 w_1'' = w_1$. 
Thus, for $x\gg 1$, we obtain the algebraic decay,   
\be 
w_1\cong G_1 x^{-\zeta},
\en 
where  $\zeta(\zeta+1)=1$ 
so $\zeta=(\sqrt{5}-1)/2$.  This  then leads to Eq.(38).

W also examine  Eq.(B1) 
in the range  $0<x<1$ because of Eqs.(36) and (37). 
In terms of  $t= \ln (1/x)>0$, 
we rewrite it   as 
\be 
{\ddot w}+ {\dot w}= -\frac{1}{4}(w-w^5)= -\frac{\p}{\p w}U(w).
\en 
We may regard $w(t)$ as a position of a particle, 
where  ${\dot w}= dw/dt$ is its velocity 
and  ${\ddot w}= d^2w/dt^2$ is its acceleration. 
Then  $\dot w$ in the left hand side of Eq.(B3) 
is a friction term.
The $U(w) $ is its  potential of the form, 
\be 
U(w)=w^2/8- w^6/24.
\en 
This potential has a local minimum at $w=0$, 
a maximum at $w=1$, and decays to negative values  for $w\gg 1$. 

First, if the initial value $w(0)$ at $t=0$ is 
smaller than 1, it decreases to 0  
 obeying  ${\ddot w}+ {\dot w}\cong -w/4$ 
for  $t\gg 1$. 
This final decay is overdamped. Thus, for $x\ll 1$, 
we find  
\be 
w \cong (A_1+ A_2 t)  e^{-t/2}, 
\en 
where $A_1$ and $A_2$ are  constants. This leads to Eq.(41). 
Second, if $w(0)>1$, $w(t)$  grows rapidly 
obeying ${\ddot w}\cong w^5/4$. Solving this 
equation yields  an explosive solution, 
\be 
w\cong 3^{1/4} /(t_m- t)^{1/2},  
\en  
where  $t_m$ is a  maximum time. This leads to Eq.(40) with  
 $x_0= e^{-t_m}<1$.  
The behaviors  (B5) and (B6) excellently agree with 
results from numerical calculations.


\begin{references}
\bibitem{Okacomment} After the renormalization of 
 the fluctuations with wave numbers 
larger than $\xi_{\rm B}^{-1}$,  
the fourth order term in the  free energy density 
becomes $C_0^2 u^* \xi_{\rm B}^{-\epsilon}\psi^4/4$ 
in the $\epsilon$ method \cite{Onukibook}, 
where $u^*$ is a universal number. 
 Some calculations give $A_c= 2^d 6^{-\epsilon/2} /u^*= 18/\pi^2\epsilon$ 
to first order in $\epsilon$ \cite{Oka}.
\bibitem{comment2} With respect to 
the scaling transformation (36), 
$x_0$ in Eq.(40) is changed to $x_0'=x_0/\lambda$. 
Also $A_1$ and $A_2$ in Eq.(41) are changed to 
$A_1'= \lambda^{1/2}(A_1 -A_2\ln \lambda)$ and 
$A_2'= \lambda^{1/2}A_2$, while 
$A_3$ in Eq.(42) is unchanged.
\end{references}
\end{document}